\documentclass[sigconf]{acmart}

\let\proof\relax

\RequirePackage{amsmath}
\RequirePackage{amssymb}
\RequirePackage{amsthm}
\usepackage{mathrsfs}
\RequirePackage{bm} 
\RequirePackage{url}
\usepackage{natbib}
\usepackage{multirow}
\usepackage{graphicx}
\usepackage{makecell}
\usepackage{booktabs}
\usepackage{array}
\usepackage{url}
\usepackage{algorithm}
\usepackage{algorithmic}
\usepackage{dsfont}
\usepackage{paralist}
\usepackage{color}
\usepackage{smile}
\usepackage{enumitem}
\usepackage{balance}

\usepackage{changepage}

\usepackage{footnote}
\makesavenoteenv{tabular}
\makesavenoteenv{table}

\usepackage{color,hyperref}
\definecolor{darkblue}{rgb}{0.0,0.0,0.3}
\hypersetup{colorlinks,breaklinks,linkcolor=darkblue,urlcolor=darkblue,anchorcolor=darkblue,citecolor=darkblue}

\usepackage{wrapfig}

\usepackage{caption}
\usepackage{subcaption}
\usepackage{environ}

\NewEnviron{myequation}{%
\begingroup
\everymath{\small}
\begin{equation}
\BODY
\end{equation}
\endgroup
}

\NewEnviron{myequation*}{%
\begingroup
\everymath{\small}
\begin{equation*}
\BODY
\end{equation*}
\endgroup
}

\NewEnviron{mycequation*}{%
\begingroup
\everymath{\small}
\begin{equation*}
\textstyle\BODY
\end{equation*}
\endgroup
}

\newcommand{\tvb}{\tilde{\vb}}

\ifx\theorem\undefined
\theoremstyle{plain}
\newtheorem{theorem}{Theorem}
\fi
\ifx\assumption\undefined
\theoremstyle{plain}
\newtheorem{thm}{Theorem}
\newtheorem{assumption}[thm]{Assumption}
\fi
\ifx\proposition\undefined
\theoremstyle{plain}
\newtheorem{proposition}[theorem]{Proposition}
\fi

\ifx\corollary\undefined
\theoremstyle{plain}
\newtheorem{corollary}[theorem]{Corollary}
\fi
\ifx\lemma\undefined
\theoremstyle{plain}
\newtheorem{lemma}[theorem]{Lemma}
\fi

\setcopyright{none}
\begin{document}
\renewcommand\footnotetextcopyrightpermission[1]{}
\pagestyle{plain}  

%
\title{Expert Finding in Heterogeneous Bibliographic Networks with Locally-trained Embeddings }
\author{
Huan Gui\footnotemark[2]\footnotemark[1]\ \ \ \ 
Qi Zhu\footnotemark[3]\footnotemark[1]\ \ \ \ 
Liyuan Liu\footnotemark[4]\ \ \ \ 
Aston Zhang\footnotemark[5]\ \ \ \ 
Jiawei Han\footnotemark[6]\\
}

\begin{abstract}
Expert finding is an important task in both industry and academia.
It is challenging to rank candidates with appropriate expertise for  various queries. 
In addition, different types of objects interact with one another, which naturally forms heterogeneous information networks.
We study the  task of expert finding in heterogeneous bibliographical networks based on two aspects: textual content analysis and authority ranking. 
Regarding the textual content analysis, we propose a new method for query expansion via locally-trained embedding learning with concept hierarchy as guidance, which is particularly tailored for specific queries with narrow semantic meanings. 
Compared with global embedding learning, 
 locally-trained embedding learning projects the terms  into a latent semantic space \emph{constrained on  relevant topics},
therefore it preserves more precise and subtle information for specific queries.
Considering the candidate ranking, the heterogeneous information network structure, while being largely ignored in the previous studies of expert finding,  provides additional information. 
Specifically, different types of interactions among objects play different roles. 
We propose a ranking algorithm to estimate the authority of objects in the network, treating each strongly-typed edge type individually.
To demonstrate the effectiveness of the proposed framework, we apply the proposed method to a large-scale  bibliographical dataset with over two million entries and one million researcher candidates. 
The experiment results show that the proposed framework outperforms existing methods for both general and specific queries.

\end{abstract}

\fancyhead{}
\settopmatter{printacmref=false, printfolios=false}

\maketitle 
{
\renewcommand{\thefootnote}{\fnsymbol{footnote}}
\footnotetext[1]{Equal Contribution}
\footnotetext[2]{huangui@fb.com, Facebook}
\footnotetext[3]{qiz3@illinois.edu, UIUC.}
\footnotetext[4]{ll2@illinois.edu, UIUC.}
\footnotetext[5]{astonz@amazon.com, Amazon AI.}
\footnotetext[6]{hanj@illinois.edu, UIUC.}
}

\setlength{\textfloatsep}{6pt}

\section{Introduction}
For a project on ``information extraction", who would be able to provide guidelines for problem solving?
For a new funding proposal on ``ontology alignment", who would be able to review and make good assessment? 
For the upcoming PKDD conference on ``data mining", who should be invited to give a keynote speech? 
\emph{Experts.}

Expert finding~\cite{balog2012expertise, deng2008formal, wang2013expertrank, zhang2007expert} is defined as the problem of ranking the candidates with appropriate expertise for a given query. 
The problem receives increasing attention in academia due to the TREC Expert Finding Track ~\cite{soboroff2006overview}.
Accurate candidate ranking has broad applications. 
However, the problem is particularly challenging since a query can be as \emph{general} as ``data mining" and ``planning" and as \emph{specific} as ``ontology alignment" and ``information extraction".
Such discrepancy among given queries poses particular challenges for accurate expert identification.  

Previous studies usually formulate the problem of expert finding as a document search problem in the information retrieval community.
Although promising results are obtained~\cite{hertzum2000information} by standard document search algorithms, the returned results are documents, not candidates. 
We take a social website as an example. 
Users actively participate in various online activities, such as posting, commenting, tagging, rating, and reviewing.  The online textual information provides evidence for users' skills and expertise.  
Moreover, users engage in online communities, collaborating, and exchanging information with each other. 
Each user cannot be simply represented by her posts or comments and she has much more complicated personal, social, and collaborative practices~\cite{deng2012modeling}.


Many approaches have been proposed and studied for expert finding. The most popular models are document-based generative probabilistic models~\cite{balog2006formal, balog2012expertise, fang2007probabilistic}. 
The major idea of the document-based models is that the expertise of a candidate can be estimated by aggregating textual evidence from relevant documents, 
which is retrieved by statistical language models. 
Nevertheless, this method suffers from the following two drawbacks. On one hand, when applying the statistical language model, there is a vocabulary gap between terms in the query and the documents.
On the other hand, such a method ignores
network structure; that is the relationships among the candidates and other objects in the heterogeneous information network. 

We attempt to solve the problem of expert finding, particularly focusing on specific queries with narrow semantic meanings without downgrading the accuracy for general queries. 
We propose a novel framework based on query expansion. It includes two different components, one is textual analysis to provide evidence  for expertise identification, and the other is authority ranking  to rank the candidates in the heterogeneous bibliographical networks. 



\noindent \textbf{Locally-trained Embedding Learning via Concept Hierarchy.} 

\begin{figure*}[!t]
\centering
\hspace*{-15mm}
\includegraphics[scale=0.27]{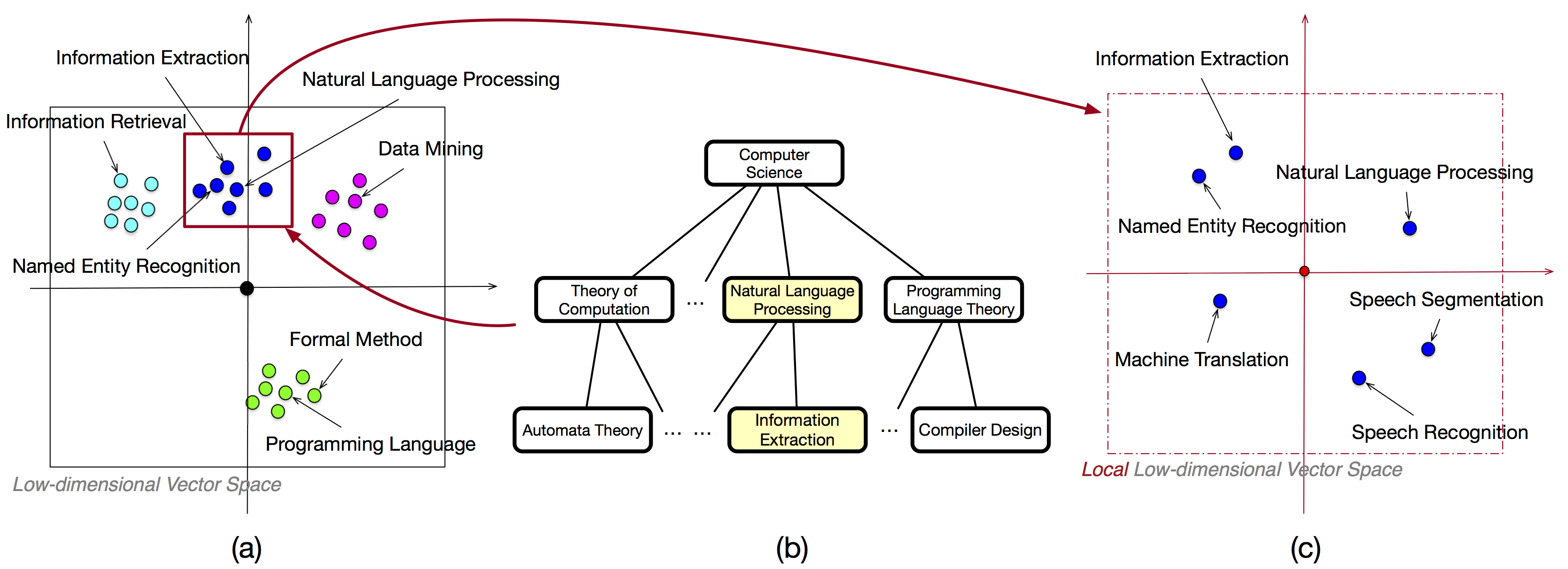}
\caption{A toy example of locally-trained embedding learning with a concept hierarchy as guidance.}\label{fig:hierarchy}
\end{figure*}

In order to address the vocabulary gap,  representation learning~\cite{cai2011graph, mikolov2013distributed, van2016unsupervised} is proposed to project the terms into a latent semantic space, such that terms with similar semantic meanings are close to each other in the latent vector space.  The vector representations  are also known as embeddings or distributed representations.
The learned embeddings are based on the co-occurrence statistics derived from the whole corpus, which can be (loosely) interpreted as a low-rank approximation for the observation data in the corpus~\cite{cai2011graph, deerwester1990indexing, levy2014neural}. 

Nevertheless, information regarding some specific queries might be missing through the semantic matching method. 
We have a toy example shown in Figure~\ref{fig:hierarchy}(a), where terms related to different domains form different clusters, such as ``information retrieval", ``natural language processing", ``data mining", and ``programming language". 
Meanwhile,  ``information extraction" is close to both ``natural language processing" and ``named entity recognition". 
Particularly for the task of expert finding, if we expand the query ``information extraction" to ``natural language processing", there will be semantic drift.

In order to address the semantic drift discussed above, we propose to train a local embedding with concept hierarchy as guidance, as shown in Figure~\ref{fig:hierarchy}(b).  
For the query ``information extraction", the \emph{cluster} that ``information extraction" belongs to can be identified as ``natural language processing". Then the local embeddings can be learned based on the documents that are relevant to ``natural language processing", as shown in Figure~\ref{fig:hierarchy}(c). 
Since the locally-trained embeddings only need to preserve the information respecting the cluster of ``natural language processing", it has stronger representation power. Consequently, the local embeddings better capture the subtle semantic information such that ``information extraction" shares closer semantic meaning with ``named entity recognition" than ``natural language processing".


\noindent \textbf{Ranking within Relevance Network.}

Extensive online {textual information} is available from candidates' activities, which serves as evidence for expertise identification.
However, the final target of expert finding is to rank \emph{candidates}, not textual information. There is a disparity. 

\begin{figure*}[!t]
\centering
\hspace*{-15mm}
\includegraphics[scale=0.37]{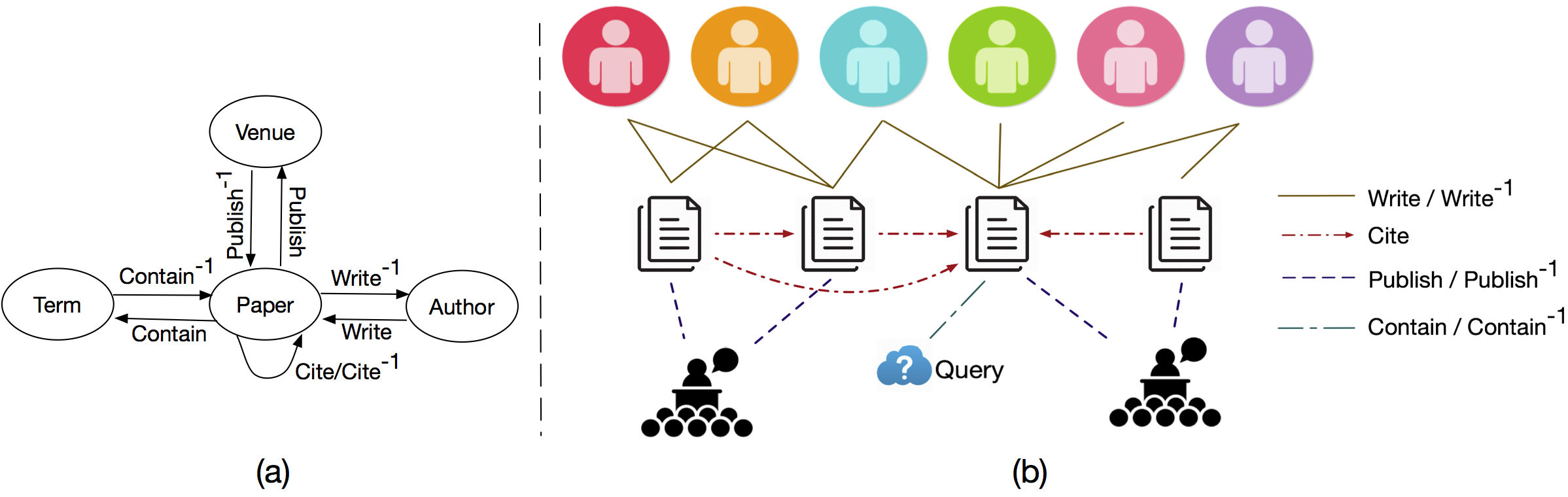}
\caption{An example heterogeneous bibliographical network with four types of objects: authors, papers, venues, and terms. (a) the network schema; (b) a subnetwork where all the documents are relevant to the  given query.}\label{fig:bibliographical}
\end{figure*}

The document-based models aggregate the relevant documents associated with each candidate and rank the candidates accordingly. 
The importance of each document is approximated by a monotonic function of the number of citations, such as logarithm functions. 
Such an aggregation method is inaccurate and sensitive to the choice of the monotonic function. 
On the other hand, besides textual information, the interactions among candidates and other objects (e.g., other candidates, group discussion in online social communities, venues in academia) offer additional insights for estimating the users' cognitive capabilities.
The interactions among the objects of different types naturally form a heterogeneous information network~\cite{sun2012mining, sun2009ranking}.
Bibliographical information network is a typical heterogeneous  information network,  which characterizes the academic publication behaviors of researchers. 
In heterogeneous bibliographical networks, researchers have various activities, including publishing, collaborating, and attending venues.
In Figure~\ref{fig:bibliographical}(a), the network schema of an example heterogeneous bibliographical network is depicted, with an illustration in Figure~\ref{fig:bibliographical}(b).

To close the gap between textual information analysis and candidate ranking, we propose a coupled random walk algorithm, including both inter-type random walks and intra-type random walks, to estimate the authority of objects in the network and the rank order of candidates.
More concretely, the ranking algorithms considers the relative importance of different edge types in the heterogeneous bibliographical network. 
%

%
%
%

To summarize, we study the problem of expert finding in heterogeneous information networks. 
Specifically, we use  the bibliographical network as a case study and the proposed framework can be straightforwardly extended to social networks and other types of networks.
The proposed framework includes two phases. The phase is locally-trained embedding learning with concept hierarchy as guidance, based on which we obtain query expansion for the given query. The second is the authority rank algorithm within the heterogeneous bibliographical network, which is retrieved and constructed based on the query expansion. 
Such a framework is particularly designed for specific queries. 
We name the new framework as \textbf{LE-expert}, which is short for \textbf{L}ocally-trained \textbf{E}mbedding for  \textbf{Expert} {F}inding.
Our contributions  are  as follows:
\begin{compactitem}
\item[1.] We propose to learn locally-trained embeddings for query expansion with a given concept hierarchy as guidance for the problem of unsupervised expert finding in heterogeneous bibliographical networks.
\item[2.] We establish a new ranking algorithm, tailored for the task of expert finding, in heterogeneous bibliographical networks.
\item[3.] We conduct numerical experiments to corroborate the efficacy of our method.  
\end{compactitem}

\section{Related Work}
In general, there are two major approaches~\cite{balog2012expertise} for the problem of expert finding, one is profile-based~\cite{liu2005finding} and the other is document-based~\cite{balog2006formal, fang2007probabilistic} (also known as the candidate and topic models). For the profile-based models, each candidate is represented via a set of terms. 
Given a query, the candidates are ranked via the ad-hoc retrieval models. 
In contrast, the document-based models are to firstly retrieve all the relevant documents  of the query and then the candidates are ranked via aggregating the associated documents. Since the document-based models make use of the whole corpus, it is usually more effective compared with the profile-based ones~\cite{balog2012expertise, deng2008formal}.
Besides these two models, there are many other approaches that take advantage of additional information. For instance, Karimzadehgan et al. propose to solve the problem of expert finding by incorporating the organizational hierarchy~\cite{karimzadehgan2009enhancing}.
The problem of vocabulary gap is addressed by query expansion with Normalized Google Distance~\cite{yang2014using}. 
More recently, an unsupervised embedding learning method is proposed, where the embeddings are learned based on the co-occurrence between candidates and terms~\cite{van2016unsupervised}. 
However, these methods mainly focus on the textual information while the  rich network structure information is ignored.

Regarding (heterogeneous) network structures, it is proposed to rank the candidates within an online forum via a propagation-based approach~\cite{zhang2007expert}. 
Besides, the problem is formalized as searching for reliable users and contents for the task of community-based query answering in a co-training fashion~\cite{bian2009learning}. 
Regarding  collaborative tagging recommendation, Noll et al. assess the expertise of users using a graph-based ranking method similar to the HITS algorithm~\cite{noll2009telling}. 
Deng et al. propose a joint optimization framework to rank candidates based on the consistency implied by the network structure~\cite{deng2012modeling}. Moreover, there are some other relevant studies, such as co-rank~\cite{zhou2007co} where authors and their publications are ranked based on a coupled random walk algorithm; NetClus~\cite{sun2009rankclus} simultaneously ranks and clusters strongly-typed objects with mutual enhancement in a heterogeneous information network; 
and RankClus~\cite{ji2011ranking} applies similar philosophy to classification and ranking. 
Nevertheless, these works are either query independent or consider the query-document relatedness based on global semantic mapping, which loses information for specific queries. Our method  not only considers the network structure, but also captures query expansion for specific queries based on locally-trained embedding learning. 

The idea of query expansion regarding local document analysis has been previously studied for information retrieval~\cite{xu1996query}. 
Global analysis and local feedback are combined for query expansion with a new weight ranking function for query expansion. 
Recently, Diaz et al. propose to perform query expansion based on locally-trained embeddings for queries with ambiguous semantic meanings~\cite{diaz2016query}.  
In contrast, our locally-trained embedding is designed for query expansion for specific queries, which is of particular importance for the task of expert finding; while theirs is for ad-hoc retrieval task~\cite{diaz2016query}. In addition, our locally-trained embeddings are learned with guidance from a concept hierarchy.
The details will be discussed in Section~\ref{sec:local-embedding}.

\section{Preliminaries}
Before detailing our method, we first introduce heterogeneous bibliographical information networks, the document-based model and word embedding learning. 
\subsection{Heterogeneous Bibliographical Networks}
A heterogeneous bibliographical network  is  constructed from bibliographical data.
Due to the heterogeneity of the object types, a heterogeneous bibliographical network is naturally a  heterogeneous information network~\cite{sun2009rankclus}. %
The formal definition of heterogeneous information networks is as follows.
\begin{definition}[Heterogeneous Information Network]
For an information network $G = (\cV, \cE)$ with an object mapping function $\phi: \cV \rightarrow \cA$ and an edge mapping function $\psi: \cE \rightarrow \cR$, where $\cA$ and $\cR$ are the set of object types and edge types, respectively, 
if the number of object types $|\cA| > 1$ or the number of edge types $|\cR| > 1$,  $G$ is  a \textit{heterogeneous information network}.
\end{definition}
DBLP  is a public bibliographical dataset in the Computer Science domain. 
We further extract semantic phrases from the text data following the method proposed by Liu et al.~\cite{liu2016representing}. Therefore, we use terms to refer both words and phrases in the corpus.
Regarding each publication entry, DBLP provides detailed information about~\emph{authors},~\emph{terms}, \emph{venues}.
Figure~\ref{fig:bibliographical}(a) depicts the network schema and Figure~\ref{fig:bibliographical}(b) is a sub-network with a user query. 
We define the set of publications as $\cD$, authors  as $\cA$, terms as $\cT$, and venues as $\cV$, with $N_D, N_A, N_T, N_V$ denoting the set sizes accordingly.

\subsection{The Document-based Models}\label{sec:slm}
The problem of expert finding has been studied extensively~\cite{balog2012expertise, deng2012modeling,  fang2007probabilistic, van2016unsupervised, zhang2007expert}.
For completeness, we present probably the most popular method: document-based models. 
The family of  document-based models formalizes the problem as a  retrieval task. 
Given a query $q$, the ranking score of a researcher candidate $a$ can be calculated as
\begin{align}\label{eq:candidate-gen}
s_c(a, q) \propto \sum\nolimits_{d \in \cD} \PP(a|d) \PP(q | d) \PP(d),
\end{align}
where  $\cD$ is the document corpus, $\PP(a|d)$ is the probability that the candidate $a$ is relevant to the publication $d$,  $\PP(q | d)$ is the probability that the query $q$ is relevant to the document $d$, and $\PP(d)$ denotes the preference over $d$.

What remains is to estimate $\PP(d), \PP(q|d)$, and $\PP(a | d)$.
Following the ideas of Deng et al.~\cite{deng2008formal}, we estimate $\PP(p)$ via  $\PP(d) \propto \ln(e + c_d)$,
where $c_d$ is the count of citations of $d$ and $e$ is the mathematical constant to guarantee that weight factor is no less than one.
$\PP(a|d)$ is generally estimated as $1/|\cA_d|$, with $\cA_d$ as the set of authors for publication $d$.
Finally,  $\PP(q|d)$ is calculated based on the query generation retrieval method with Dirichlet prior smoothing~\cite{zhai2004study}, 
\begin{align}~\label{eq:query-likelihood-model}
\PP(q | d) \propto  \exp \Big( \sum\nolimits_{t \in q} \PP(t | q) \log \PP (t | \theta_d) \Big),
\end{align} 
where $ \PP(t | q) = {\#(t, q)}/{\#(q)} $ with $ \#(t, q)$ as term frequency of term $t$ in $q$ and $\#(q)$ as the length of $q$ and $\PP (t | \theta_d) $ is defined as 
\begin{align}\label{eq:qlm}
\PP (t | \theta_d) = \beta \PP(t | d) + (1 - \beta) \PP_{\rm b}(t),
\end{align}
with $\beta = 0.5$ and $\PP_{\rm b}(t)$ as the background language model of the text corpus $\cD$.

\subsection{Word Embedding Learning}\label{sec:embedding}
Word embedding learning~\cite{mikolov2013distributed, pennington2014glove} is to represent the terms in a corpus into a low-dimensional  latent semantic space,
where each term is represented via a low-dimensional vector, which is  called embedding or distributed representation.
The semantic information regarding each term is preserved such that terms with similar semantic meanings are close to each other in the Euclidean space.
%
There are many off-the-shelf embedding learning algorithms. 
We adopt word2vec~\cite{mikolov2013distributed} to learn the embeddings, and other embedding methods, such as Latent Semantic Indexing and Glove~\cite{pennington2014glove}, can also be applied. 
In word2vec, for a pair of words that co-occur in a sliding window, one term $u$ is denoted as target and the other $v$ as context. 
Based on the skip-gram model, 
the conditional probability of observing $u$ given $c$ is defined using the softmax function
\begin{align}\label{eq:bipartite-cond}
\PP(u | c)  = \frac{\exp(\vb_u ^\top \tvb_c)}{\sum_{u' \in \cT} \exp(\vb_{u'}^\top \tvb_c)},
\end{align}  
where $\vb_u,  \tvb_c \in \RR^z$ are the embeddings for $u~\text{and}~c$, with $z$ as the dimension of the embedding vector.
In~\eqref{eq:bipartite-cond}, since the denominator sums over all the terms in the corpus $\cT$, it is computationally intractable. Consequently, negative sampling is proposed~\cite{mikolov2013distributed}. 
For the term pair of $(u, c)$, regarding~\eqref{eq:bipartite-cond}, the following objective is optimized instead,
\begin{align}\label{eq:obj-1}
\ell(u,  c) = \log \sigma(\vb_u ^\top \tvb_c) + \sum\nolimits_{i = 1}^g \EE_{u_n \sim P_n } \big[ \log \sigma( - \vb_{u_n} ^\top \tvb_c) \big],
\end{align} 
where $\sigma(\cdot)$  is the sigmoid function, $ \vb_{u_n} \in \RR^d$ is the embeddings of noise $ u_n$, $P_n$ is the noise term distribution, and $g$ is the negative sampling parameter. 
Due to space limit, one may refer to the original paper by Mikolov et al. for technical details~\cite{mikolov2013distributed}.

\section{Local Embedding via Concept Hierarchy}\label{sec:local-embedding}
Word embedding learning is proposed for \emph{global} embedding learning such that an embedding vector is learned for each term regarding the whole corpus. 
According to Levy et al.~\cite{levy2014neural}, the word embedding learning with negative sampling in~\eqref{eq:obj-1} can be loosely interpreted as  an implicit matrix factorization problem, where the shifted positive Pointwise Mutual Information (PMI) matrix is approximated by a low-rank matrix with rank  equivalent to $z$ (the dimension of the vector space). 
However, such an approximation may lead to coarse representations of specific terms. 
The term ``information extraction" is not only close to ``information extraction'' and ``named entity recognition" but also to ``text mining" and ``natural language processing". 
Suppose that ``natural language processing" was used as expansion of ``information extraction", there will be a semantic drift. 
Instead of obtaining experts on ``information extraction" only, we may also find experts on ``natural language processing".
However, not all of the experts on ``natural language processing" are working on  ``information extraction".
%
\begin{table*}[!t]
\caption{List of terms most similar to ``Information Extraction" by different embedding methods: (i) global embedding; (2)  the method proposed by Diaz et al.~\cite{diaz2016query} without a concept hierarchy, denoted as LE wo/ CH (LE: locally-trained emebddings, CH: concept hierarchy);  (3) the proposed locally-trained embedding learning with a concept hierarchy as guidance, denoted as LE w/ CH. } \label{tab:info-ext}
\centering
\hspace*{-3mm}
\begin{tabular}{|c|c|c|}
\hline
Global Embedding & LE wo/ CH~\cite{diaz2016query}  & LE w/ CH \\ \hline
information-extraction-ie &  pattern-discovery  & information-extraction-ie\\ \hline
text-mining & knowledge-based & SystemT~\footnote{IBM SystemT is a declarative information extraction system.
 } \\ \hline
natural-language-processing & indices  &  ontology-based-information-extraction\\\hline
question-answering & legal &  web-information-extraction\\ \hline
named-entity-recognition & turkish  & relation-extraction\\ \hline
nlp & offer  & named-entity-recognition \\ \hline \end{tabular}
\vspace*{-1mm}
\end{table*}

\subsection{Concept Hierarchy}
In order to address the semantic drift, we relax the global low-rank assumption and propose to represent the terms in the corpus using locally-trained embeddings. In particular, we make the following assumption.
\begin{assumption}
The shifted positive PMI matrix is low-rank {for a sub-corpus that is relevant to the query}.
\end{assumption}
The sub-corpus is constructed with guidance from a concept hierarchy (Figure~\ref{fig:hierarchy}(b)).
In other words, instead of learning embeddings to preserve the information in the whole corpus, we only preserve information in the sub-corpus. 
The sub-corpus corresponds to the cluster that ``information extraction" belongs to, which is ``natural language processing" according to the concept hierarchy in Figure~~\ref{fig:hierarchy}(b).
Therefore, the sub-corpus comprises publication documents constrained on ``natural language processing".

 \emph{Why using  a concept hierarchy as guidance?}
Regarding the task of expert finding, for a given query ``information extraction", the (implicit)  background information is  ``natural language processing". 
By taking advantage of  concept hierarchy, we can identify the background information, as depicted in Figure~\ref{fig:hierarchy}(b). 
Alternatively, without a concept hierarchy, as proposed by Diaz et al.~\cite{diaz2016query}, the sub-corpus is constructed by retrieving all the documents relevant to the ``information extraction". 
 The results obtained following the idea of Diaz et al.~\cite{diaz2016query} are shown in the second column of Table~\ref{tab:info-ext}.
However, the top-ranked terms are random and irrelevant to ``information extraction".
This is because when learning term embeddings on sub-corpus constrained on ``information extraction", the term  ``information extraction" becomes the background since it appears in almost all the documents and (almost) co-occur with all words in the corpus, especially for short documents.
In the bibliographical data that we use, around 76\% of the document entries are titles. Therefore, ``information extraction" is similar to stop words.
Meanwhile, if the sub-corpus is constrained on ``natural language processing", the term ``natural language processing" becomes the background and is distant from ``information extraction", as shown in the third column of Table~\ref{tab:info-ext}.

\subsection{Locally-trained Embedding Learning}\label{sec:llel}
\emph{How to use  concept hierarchy as guidance for local embedding learning?}
For brevity, we first consider the case where there is only one term in each query, corresponding to one concept in the concept hierarchy. 
Also, we assume that terms in the query can be trivially mapped to  the concept hierarchy.
For queries with more than one concept, we train local embeddings one by one. 
For each concept, we use the learned local embeddings to expand the concept accordingly.


For a given query $q$ in the concept hierarchy, we denote the path from root to $q$ as  $\cC_{0} \rightarrow \cC_{1} \rightarrow \ldots \rightarrow \cC_{l} = q$, where $l$ is the level of the concept hierarchy that $q$ lies at and $\cC_{0}$ corresponds to the root. 
We use $\big\{ \vb_{t}^m \big \}_{t \in \cT}$ for $m = 0, \ldots, l$ to denote the learned embeddings for terms at level $m$.
The idea of local embedding learning is to \emph{find the nearest neighbors  (i.e., expansions) of $\cC_m$ based on the term embeddings learned constrained on $\cC_{m - 1}$}. 
Therefore, the nearest neighbors of $q$ can be found based on the embeddings learned on a sub-corpus constrained on $\cC_{l - 1}$.
In the following, we use ``information extraction" as a running example. 

For the (sub-)corpus constrained on concept $\cC_0$, it is straightforward that we use the whole corpus to train terms' embeddings (i.e., global embeddings). 
For the corpus constrained on concept $\cC_m$ for $m = 1, 2, \ldots$, we first search for the $k$ nearest neighbors of $\cC_m$, which serve as expansions to close the vocabulary gap while constructing the sub-corpus.  For the query ``information extraction", we have $\cC_1 = $ ``natural language processing".

As we do not have features for each concept (and term), we use the embeddings learned via a sub-corpus constrained on concept $\cC_{m - 1}$ as features. 
Given $\big\{ \vb_t^{m - 1} \big \}_{t \in \cT}$ as the embedding learned constrained on $\cC_{m - 1}$, we use cosine similarity to measure the similarity between term $s_{m - 1}(t_1, t_2)$.
The top $k$ terms measured by $s_{m - 1}(\cdot, \cC_{m})$ is denoted $\cN_{m}$, as expansion of concept $\cC_{m}$.
Therefore, a sub-corpus constrained on $\cC_m$ can be extracted based on $\cN_{m}$. 
In other words, we use global embeddings ($\big\{ \vb_t^0 \big \}_{t \in \cT}$) to firstly find the query expansions of ``natural language processing", which is denoted as $\cN_1 = \{$ ``natural language processing", ``nlp",  ``natural language understanding", ``language processing", $ \ldots \}$. 
We interpolate such semantic similarity into the language model with parameter $\gamma \in [0, 1]$, 
\begin{align}\label{eq:qlm2}
\PP^{m}(t | d) =  \gamma \PP(t | d) + (1 - \gamma) s_{m - 1}(t, \cC_{m}) \Ib(t \in \cN_{m}), 
\end{align}
where $\Ib(w \in \cN_{m})$ is an indicator function.
 Substituting~\eqref{eq:qlm2} into~\eqref{eq:qlm} and~\eqref{eq:query-likelihood-model}, we obtain $\PP^{m}(q | d)$:
\begin{align} \label{eq:sample}
\PP^{m}(q | d) = \PP^{m}(t | d)  =  \beta \PP^{m}(t | d) + (1 - \beta) \PP_b(t),
\end{align} 
where query $q =\{t\}$ contains only one concept, $t$. 
In order to train local embeddings on the sub-corpus constrained on $\cC^{m-1}$, we sample each document with probability proportional  to $\PP^{m-1}(q | d)$. We set $\PP^{0}(q | d) = 1 / |\cD|$, as the uniform sampling. 
While applying word2vec for embedding learning, in order to estimate  the empirical distribution of terms in the sub-corpus constrained on $\cC_m$, the sampling weights of each document (i.e.,  $\PP^{m}(q|d)$) should be considered.
The recursive embedding learning  framework is detailed in Algorithm~\ref{algo:lel}. 
\begin{algorithm}[!t]
	\caption{{Local Embedding Learning via Concept Hierarchy}.}\label{algo:lel}
	\begin{algorithmic}[1]
		 \STATE \textbf{Input:} Document corpus $\cD$, the path to query as $\cC_0 \rightarrow \cC_1 \rightarrow \ldots \rightarrow \cC_l  = q$.
		 \STATE \textbf{Initialize:} $\cS = \cD$. 
		 \FOR{$m = 0, \ldots, l - 1$}
		 \STATE Learn embeddings of $t \in \cT$ using word2vec
		 \STATE ~~~~ Sample each document with probability $\propto \PP^{m}(q|d)$
		 \STATE Output $\vb_t^{m}$ as the embeddings of term $t$ 
		 \STATE Compute $\PP^{m + 1}(q|d)$ according to~\eqref{eq:sample}.
		 \ENDFOR
		\STATE \textbf{Return:} $\vb_t^{l}$. 
	\end{algorithmic}
\end{algorithm}

\section{Expert Ranking in Relevance Network}\label{sec:ranking}
In order to rank researcher candidates for each query, we have two key insights:
(i) A candidate may have papers on many topics. For a given query, only the relevant papers can serve as textual evidence for expertise. 
(ii)  Citation may have time-delay factor. Papers that are published in a higher-ranked venue are more likely to be important. Therefore, venues play an important role for ranking.
\subsection{Relevance Network Construction}
For a given query $q$, we first retrieve all the relevant documents,  the set of which is denoted as  $\cD(q) = \big \{ d : \prod\nolimits_{t_i \in q} \max\nolimits_{t' \in \tilde{\cN}_l(t_i)} \big \{ \Ib(t' \in d) \big \} = 1 \big  \}$, where $\Ib(t' \in d) = 1$ if $t'$ is within the document $d$, $0$ otherwise. 
In other words, we select all the papers that contain at least one relevant term in $\tilde{\cN}_l(t_i)$ for each term $t_i~(i = 1, 2, \ldots, N_q)$ in $q$. 

Based on $\cD(q)$, a relevance sub-network can be extracted from the heterogeneous bibliographical network by extracting $\cD(q)$ and associated authors and venues. 
LE-expert ranks the candidates within the  relevance sub-network. 


\subsection{Ranking in Relevance Network}
To rank candidates for each query, we take advantage of the network structure and propose
a ranking algorithm to estimate the authority of objects in the sub-network based on a coupled random walk in the relevance sub-network.
We first present the ranking method in a general framework, which can be generalized for other heterogeneous information networks. 

Suppose there are $M$ types of objects in the heterogeneous information network and the set of the type $i$  objects is denoted as $\cV_i$.
The  network is represented by a set of relation matrices $\cR = \{ \Rb_{ij} \}_{i, j = 1}^M$.
For each $\Rb_{ij}$, we define a diagonal matrix $\Db_{ii}$ such that the diagonal element at $(a, a)$ of $\Db_{ij}$ is the sum of the $a$-th row of $\Rb_{ij}$. Therefore,  the transition matrix of $\Rb_{ij}$ is defined as $\Pb_{ij} = \Db_{ij}^{-1}\Rb_{ij}$. 
And the ranking score vector of objects in type $i$  can be updated iteratively:
\begin{align}\label{eq:iter-ranking}
\rb_{i}^{t} \propto \frac{\sum_{j = 1}^M \lambda_{ji} \rb_{j}^{t - 1} \Pb_{ji} + \eta_i \rb_{i}^{0}    }{ \sum_{j = 1}^M \lambda_{ji} + \eta_i},
\vspace*{-2mm}
\end{align}
where  $t$ is the iteration step and $\rb_{i}^{0} = \mathbf{1} / |\cV_i|$. 
The relative importance of neighbors of different types is controlled by $\lambda_{ij} \in [0, 1]$. 

Regarding the task of expert finding in heterogeneous bibliographical networks, the random rank is designed regarding the following assumption~\footnote{Since terms are used to construct the relevance network and do not reflect authority, we do not consider  terms while ranking the candidates.}.
\begin{assumption}\label{ass:ranking}
\begin{itemize}
\item[(a)] High-quality and relevant papers will be frequently cited by many other relevant papers;
\item[(b)] Relevant highly-ranked experts will publish many high-quality and relevant papers, and vice versa. 
\item[(c)] Relevant and highly-ranked conferences attract many high-quality and relevant papers, and vice versa. 
\end{itemize}
\end{assumption}
Therefore, the relation types for each object type are as follows.
 \textbf{Paper:}  (i) Citation relations. $\Rb_{\rm PP}(a, b) = 1$ if the  paper $a$ cites the   paper $b$; (ii) Write relations. $\Rb_{\rm AP}(a, b) = 1$ if  author $a$ writes  paper $b$; (iii) Publish relations. $\Rb_{\rm VP}(a, b) = 1$ if  paper $b$ is published in venue $a$. 
  \textbf{Author:}  (i) Coauthor relations. $\Rb_{\rm AA} = \Rb_{\rm AP} \Rb_{\rm AP}^\top$; (ii) Write$^{-1}$ relations. $\Rb_{\rm PA} = \Rb_{\rm AP}^{\top}$.
 \textbf{Venue:} (i) Citation relations.  $\Rb_{\rm VV} = \Rb_{\rm VP}  \Rb_{\rm PP} \Rb_{\rm VP}^\top$ (ii) Publish$^{-1}$ relations. $\Rb_{\rm PV} = \Rb_{\rm VP}^{\top}$.

\begin{remark}
The underlying philosophy  of the ranking module is similar to NetClus~\cite{sun2009rankclus} and RankClass~\cite{ji2011ranking}.
However, NetClus and RankClass are primarily designed for clustering and classification in the whole heterogeneous information network, respectively; while  
 LE-expert is designed for authority ranking within a relevance sub-network.
In addition, NetClus can only be applied to star-schema heterogeneous  networks while LE-expert is independent of the network schema.
Moreover, RankClass is a regularization framework for label propagation whereas  LE-expert is based on  random walks. 

\end{remark}

\section{Experimental Results}
We conduct various  experiments to study effectiveness of the proposed framework in expert finding for both specific and general queries.
\subsection{Experimental Setup}
\textbf{Data.} To evaluate the proposed framework, we conduct numerical experiments and case studies on the dataset of DBLP. 
In the DBLP dataset, there are 2,244,018 papers, 1,274,360 authors, 8,882 venues, and  1,812,277 words and phrases. Among all the papers, 529,498 papers (24\%) have abstract information. 
The labelled dataset is from Deng et al.~\cite{deng2012modeling}, which contains 20 queries in total, including both general and specific ones. Details on the queries and the number of experts for each query can be found in~\cite{deng2012modeling}.

\noindent \textbf{Evaluation Measures.} Regarding evaluation of the task, we employ several popular information retrieval measures~\cite{buttcher2016information}, including Precision at rank $n $ (P$@n$), Mean Averaged Precision (MAP),   Normalized Discounted Cumulative Gain at rank $n$ (NDCG$@n$), and bpref~\cite{buckley2004retrieval}. 
P$@n$ measures the percentage of relevant experts in the top $n$ of the retrieved candidate list, which is estimated as P$@n = \sum_{i = 1}^n R(c_i) / n$, where $R(c_i) = 1$ if the $i$-th retrieved candidate is relevant to the given query and $R(c_i) = 0$ otherwise.  Suppose there are $R_n$ relevant experts, Average Precision is defined as AP$= \sum_{i = 1}^{R_n} ({\rm P}@i * R(c_i)) / R_n$ and MAP is the averaged AP for all queries. Since the relevance labels are binary, therefore, NDCG is defined as NDCG$@n = \sum_{i = 1}^n R(c_i) / \log_2(i + 1) / \sum_{i =1 }^n \big[ 1 / \log_2(i + 1) \big]$.
Also, we consider bpref, which is a summation based measure of the number of relevant documents ranking before irrelevant ones, bpref $=R_n^{-1}\sum_{r=1}^{R_n}(1 - \sum_{i = 1}^r (1 - R(c_i)) / R_n).$

\noindent  \textbf{Baselines.}  We compare LE-expert with the following baselines:
\begin{compactitem}
\item \textsc{Balog}. It is a classical  document-based model for expert finding~\cite{balog2006formal}.
\item NMF. We apply nonnegative matrix factorization~\cite{cai2011graph} to the author-term co-occurrence matrix. The ranking of authors is based on the inner product of the corresponding rows and columns of authors and queries. 
\item LSI. We apply latent semantic indexing to identify the similarity of the authors and the queries. 
\item \textsc{Corank~\cite{zhou2007co}.} Co-ranking cannot be directly applied for expert finding since it is query independent. Therefore, we first retrieve relevant documents and then apply co-ranking for each query. 
\item {Embed}~\cite{van2016unsupervised}. This is a global embedding algorithm designed for the task of unsupervised expert finding without considering the network structure. 
\item JointHyp~\cite{deng2012modeling}. JointHyp is a regularization framework for expert finding in heterogeneous information networks. Specifically,  information is propagated through the network based on consistency in the network.
\item Exact. The relevance sub-network is extracted based on the exact match.
\item RankClass. The sub-relevance sub-network is extracted based on query expansion, and rank the candidates by RankClass with only one class. 
\end{compactitem}
For fair comparison, we use the same leave-one-out cross-validation dataset and report the best performance of each model. 
The parameter setting of  LE-expert   is as follows $\beta = \gamma = 0.5$ in~\eqref{eq:qlm2} and~\eqref{eq:sample}.  We gradually reduce the size of dimension of the local vector space and set $z = \lceil 300 / (5m + 1) \rceil$ with $m$ being the hierarchy level. 
For the concepts $\cC_m$, the size of the query expansion ($\cN_m$) is set to be $k = 30$. 
The final expansion for each query ($\tilde{\cN}_m$) is set by cross validation. 
Recall that $\tilde{\cN}_m$ is query expansion set for relevance sub-network construction. It is worth noting that general queries are more likely to have more expansions and specific ones have less expansions.
\begin{table*}[!t]
\caption{Overall evaluation results.} \label{tab:result-2}
\centering
\begin{tabular}{@{}|c|c|c|c|c|c|c|c|c|@{}} \hline
measure & P@5 & P@10 & P@20 & NDCG@5 & NDCG@10 & NDCG@20 & MAP & bpref\\ \hline \hline  
\textsc{Balog} & 0.4941 & 0.3824 & 0.2853 & 0.5068 & 0.4248 & 0.3416 & 0.1608 & 0.8536 \\ \hline 
\textsc{NMF} & 0.3176 & 0.2706 & 0.2118 & 0.3525 & 0.3075 & 0.253 & 0.1151 & 0.7303  \\ \hline 
\textsc{SVD} & 0.4353 & 0.3471 & 0.2912 & 0.4553 & 0.3871 & 0.3336 & 0.1548 & 0.7590 \\ \hline 
\textsc{Corank} & 0.6941 & 0.5741 & 0.4235 & 0.7181 & 0.6386 & 0.5024 & 0.291 & 0.8843 \\  \hline 
\textsc{Embed} & 0.0353 & 0.0294 & 0.0265 & 0.0354 & 0.0317 & 0.0289 & 0.005 & 0.6331 \\  \hline 
\textsc{JointHyp}  & 0.6235 &  0.4176 &  0.2882 &  0.6447 & 0.4913  &  0.3725 & 0.1579  & \boldmath\textbf{0.9704} \\ \hline 
\textsc{Exact} & 0.7059 & 0.5882 & 0.4529 & 0.7548 & 0.6549 & 0.5361 & 0.311 & 0.8676  \\  \hline 
RankClass & 0.7529 & 0.6647 & 0.5176 & 0.7666 & 0.7026 & 0.5867 & 0.3598 & {0.8981}\\ \hline
LE-expert & \boldmath\textbf{0.8118} & \boldmath\textbf{0.7118} & \boldmath\textbf{0.5559} & \boldmath\textbf{0.8027} & \boldmath\textbf{0.7361} & \boldmath\textbf{0.618} & \boldmath\textbf{0.3826} & {0.9451} \\ \hline
\end{tabular}
\end{table*}

\subsection{Experimental Results}

\noindent \textbf{Overall Results Analysis.} 
The experimental results of different methods are summarized in Table~\ref{tab:result-2}.
%
Compared with \textsc{Balog, NMF, SVD}, and \textsc{Embed} which only  utilize  the textual information and the overall number of citations  as the prior of each document, as shown in Table~\ref{tab:result-2}, we can see that the methods that take advantage of the network information, including \textsc{Corank, JointHyp, Exact}, RankClass, and LE-expert, achieve significantly better results regarding all the evaluation measures. 
This result agrees with our argument that the task of expert finding is different from information retrieval and the network structure plays an important role. 
Moreover, we notice that the precision of Embed~\cite{van2016unsupervised} is even worse than classical embedding methods, such as NMF and SVD. It can be partially explained by that for a candidate $c$ and a query $q$ the ranking score  can be (loosely) interpreted as scaling with $\#(c, q) / (\#c \#q)$~\cite{levy2014neural}, which favors candidates with more focused expertise.
More specifically, a candidate with only one paper on $q$ is likely to be ranked topmost.

Now we consider the methods taking advantage of the heterogeneous network structure.
Comparing \textsc{Corank} with \textsc{Exact}, we see that \textsc{Exact} performs slightly better than \textsc{Corank} in measures of Precisions, NDCG's, and MAP. This is because \textsc{Exact} additionally considers the venue information for ranking.   
Moreover,  LE-expert significantly outperforms \textsc{Exact} regarding all the evaluation measures, which serves as evidence that the proposed query expansion method can solve the problem of vocabulary gap. 
Unlike  the global embedding methods (NMF and SVD), LE-expert will not expand specific queries to more general ones thanks to the locally-trained embeddings.
LE-expert achieves better precision and NDCG results. 
 \textsc{JointHyp}~\cite{deng2008formal} is also designed for heterogeneous bibliographical information networks, the main idea of which to propagate  the relevance of documents for each query to the candidates through the strongly-typed edges in the network. 
 However, such a method will give inaccurate estimation for documents regarding specific queries since the relevance of documents is estimated via global embeddings. 
Our model is based on the coupled random walks, where the weights for all documents are the same (as $\rb_i^0  = 1/ |\cV_i|$). The prediction accuracy of LE-expert is better than \textsc{JointHyp}; while  \textsc{JointHyp} slightly outperforms ours regarding the overall ranking (bref). 
However, it is worth noting that for the task of expert finding, the top-ranked  results are more important. 
We also compare LE-expert against RankClass, which is similar w.r.t. the ranking algorithm. 
RankClass is a regularization framework; while LE-expert considers the inter-type and intra-type random walks.
We can see that LE-expert performs better than RankClass on precision and NDCG results. 

\begin{figure}[!t]
\centering
\includegraphics[scale=0.20]{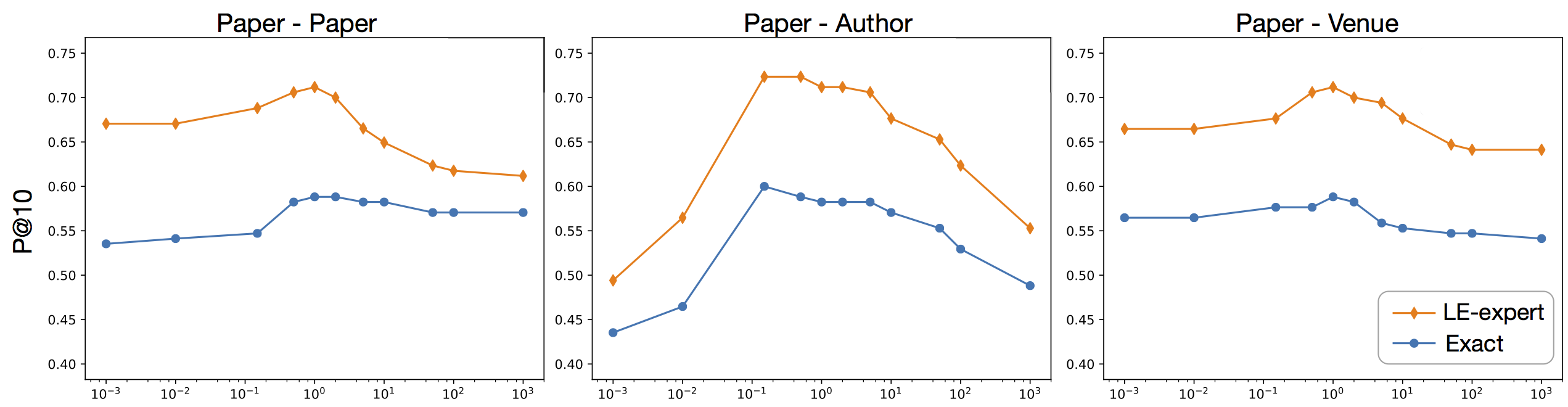}
\caption{The Precision@10 scales with the hyper parameter of the weight of different relation types.}\label{fig:Hyperparameter-2}
\vspace*{-4mm}
\end{figure}

\noindent \textbf{Hyper parameter.}
As shown in~\eqref{eq:iter-ranking},  hyper parameter $\lambda_{\cdot, \cdot}$ (the relative importance of different types of edges) plays an important role for the final ranking of candidates. 
The sensitivities of the ranking results with varying $\lambda_{\cdot, \cdot}$ is depicted in Figure~\ref{fig:Hyperparameter-2}.
For simplicity, we set $\lambda_{i, j} = \lambda_{j, i}$  for all $i, j$. In addition, except for the $\lambda$ of interest, all the other $\lambda_{\cdot, \cdot} = 1$. 
The y-axis corresponds to Precision@10. Firstly, we observe that the ranking results are more sensitive to $\lambda_{PA}$ than $\lambda_{PP}$ and $\lambda_{PV}$. This can be explained by the fact that the ranking is based on authors. 
The second observation is that the precision results follow a pattern that the precision first increases then decreases as the weight parameters increase.  
For one edge type, if the corresponding $\lambda$ goes to zero, it is equivalent to removing that edge. 
Such an observation indicates that all the edge types are involved in the final ranking. 
Our third observation is that when $\lambda_{PP}$ and $\lambda_{PV}$ go to zero, the performance remains stable; when $\lambda_{PA}$ goes to zero, the performance drops significantly. This can be explained by that $\lambda_{PA}$ balances the relative importance between coauthor relations and writing relations.
When $\lambda_{PA}$ goes to zero, the ranking order candidates is dominated by coauthor relations.
The absence of authority information from papers leads to fallacious ranking results.
Meanwhile, when $\lambda_{PA}$ goes to infinity, the ranking model is reduced to the document retrieval model (with the relevance of each document to be equal), since the other types of edges do not contribute to the authority scores of candidates. 
\begin{table*}[!t]

\centering
\small
\caption{Case study. } 
\begin{tabular}{@{}cc|cc@{}} \hline
 \multicolumn{2}{c|}{\textbf{boosting}}  &  \multicolumn{2}{c}{\textbf{support vector machine}}   \\ 
\textsc{Corank} & LE-expert & \textsc{Corank} & LE-expert \\   \hline
\boldmath\textbf{Robert E. Schapire} & \boldmath\textbf{Robert E. Schapire}  & Qi Wu & \boldmath\textbf{Bernhard Sch\"{o}lkopf} \\
\boldmath\textbf{Yoav Freund} & \boldmath\textbf{Yoav Freund} & Isabelle Guyon & \boldmath\textbf{Vladimir Vapnik} \\
Ron Kohavi & \boldmath\textbf{Leo Breiman}  & \boldmath\textbf{Jason Weston} & \boldmath\textbf{C. J. C. Burges}\\
\boldmath\textbf{Thomas G. Dietterich} & \boldmath\textbf{Yoram Singer}  & \boldmath\textbf{Vladimir Vapnik} & \boldmath\textbf{Thorsten Joachims} \\
\boldmath\textbf{Yoram Singer} & \boldmath\textbf{David P. Helmbold} & Bao-Kiang Lu & \boldmath\textbf{Chih-Jen Lin} \\ \hline \hline 
 \multicolumn{2}{c|}{\textbf{information extraction}}  & \multicolumn{2}{c}{\textbf{ontology alignment}}  \\ 
\textsc{Corank} & LE-expert & \textsc{Corank} & LE-expert \\  \hline 
\boldmath\textbf{Ralph Grishman} & \boldmath\textbf{Dayne Freitag} & \boldmath\textbf{Jerome Euzenat} & \boldmath\textbf{W. M. Schorlemmer}\\
\boldmath\textbf{Andrew McCallum} & \boldmath\textbf{Ralph Grishman} & Patrick Lambrix & \boldmath\textbf{Yannis Kalfoglou} \\
Ellen Riloff & \boldmath\textbf{Andrew McCallum} & Jason J. Jung & \boldmath\textbf{Anhai Doan} \\
Oren Etzioni & \boldmath\textbf{Nicholas Kushmerick} & He Tan & \boldmath\textbf{Jerome Euzenat}\\
\boldmath\textbf{Dayne Freitag} & \boldmath\textbf{Stephen Soderland} & Marc Ehrig & Alon Y. Halevy\\ \hline
\end{tabular}
\label{tab:case-study}
\end{table*}

\noindent \textbf {Case Study.}
Some concrete case studies of candidate ranking are shown in Table~\ref{tab:case-study}. 
For general queries, including ``boosting" and ``support vector machine",  the query expansions are based on the global embeddings. 
LE-expert has  better precision. Specifically, for ``support vector machine",  ``Bernhard Sch\"{o}lkopf", who makes particular contributions to ``support vector machine", and ``Vladimir Vapnik", who is a co-inventor of the support vector machine, rank topmost. 
This demonstrates the power of the proposed framework in general queries. 
For specific queries, we consider ``information extraction" (as a child of ``natural language processing") and ``ontology alignment" (as a child of ``ontology"). 
The high  precision results of specific queries indicate that the locally-trained embedding learning method provides accurate and relatively complete expansions for the queries. Moreover, the ranking algorithm contributes to the authority ranking of candidates. 
Taking ``information extraction" as an example, ``Dayne Freitag" whose research is focusing on ``machine learning for information extraction" ranks higher than more senior researchers ``Ralph Grishman" and ``Andrew McCallum", given that all of them work on ``information extraction".

\section{Conclusions and Future Work}
In this paper, we study the problem of expert finding in heterogeneous bibliographical information networks based on a query expansion approach. Firstly, we propose to perform query expansion based on locally-trained embedding learning recursively with a concept hierarchy as guidance. Secondly, we introduce a ranking algorithm on a relevance sub-network to estimate the expertise of the candidates via coupling both inter-type and intra-type random walks.  
Numerical experimental results on a large-scale  heterogeneous bibliographical information network corroborate the effectiveness of the proposed LE-expert.

The proposed framework is general and can be applied to other tasks, such as query-answering in online communities or recruiting for open problem solving. Besides, the locally-trained embedding learning with a concept hierarchy as guidance is of independent interest and may be applied to other tasks, such as product recommendation given a product hierarchy. 
In addition, since our framework requires a concept hierarchy as the input, we plan to consider a more challenging scenario where concept hierarchy is not available  for future work.

\bibliographystyle{abbrv}
{\scriptsize
\bibliography{reference}
}

\end{document}